\documentclass[final]{aipproc}

\def\lesssim{\mathrel{\hbox{\rlap{\hbox{\lower4pt\hbox{$\sim$}}}\hbox{$<$}}}}
\def\gtrsim{\mathrel{\hbox{\rlap{\hbox{\lower4pt\hbox{$\sim$}}}\hbox{$>$}}}}

\layoutstyle{6x9}

\begin{document}

\title{A brief survey of LISA sources and science}

\classification{04.30.Db}
\keywords{Gravitational waves}

\author{Scott A.\ Hughes}{
  address={Department of Physics and MIT Kavli Institute\\
  Massachusetts Institute of Technology, Cambridge, MA 02139} }

\begin{abstract}
LISA is a planned space-based gravitational-wave (GW) detector that
would be sensitive to waves from low-frequency sources, in the band of
roughly $(0.03 - 0.1)\,{\rm mHz} \lesssim f \lesssim 0.1\,{\rm Hz}$.
This is expected to be an extremely rich chunk of the GW spectrum ---
observing these waves will provide a unique view of dynamical
processes in astrophysics.  Here we give a quick survey of some key
LISA sources and what GWs can uniquely teach us about these sources.
Particularly noteworthy science which is highlighted here is the
potential for LISA to track the moderate to high redshift evolution of
black hole masses and spins through the measurement of GWs generated
from massive black hole binaries (which in turn form by the merger of
galaxies and protogalaxies).  Measurement of these binary black hole
waves has the potential to determine the masses and spins of the
constituent black holes with percent-level accuracy or better,
providing a unique high-precision probe of an aspect of early
structure growth.  This article is based on the ``Astrophysics
Tutorial'' talk given by the author at the Sixth International LISA
Symposium.
\end{abstract}

\maketitle

Our view of the universe today is built almost entirely by exploiting
the electromagnetic interaction\footnote{With a relatively small, but
extremely important, complementary exploitation of the weak
interaction via neutrino astronomy.}.  The leading order radiation
comes, as is very well known, from time variations of the charge
dipole moment $d_i$ of a source: writing
\begin{equation}
d_i = \int \rho_{\rm e}(x')x'_i\,d^3x'\;,
\end{equation}
(where $\rho_{\rm e}$ is the charge density and the integral is taken
over an extended source), the radiative electromagnetic 4-potential is
given by
\begin{equation}
A_i \simeq \frac{{\dot d}_i}{r}\sim \frac{vq^{\rm dip}}{r}\;.
\end{equation}
The final order of magnitude formula tells us that the amplitude of
this radiation is set by the amount of charge participating in dipole
motions, and the speed of that charge's motion.  Since many
astrophysical environments are rich in plasma and free charges,
electromagnetic radiation is readily created; it is likewise detected
with relative ease.

Gravity also generates radiation.  Since ``gravitational charge''
(i.e., mass) only comes with one sign, the leading order radiation is
quadrupolar rather than dipolar (see {\cite{fh05}} for discussion):
defining the mass quadrupole moment
\begin{equation}
Q_{ij} = \int \rho_{\rm m}(x')\left(x'_i x'_j -
\frac{1}{3}(r')^2\delta_{ij}\right)d^3x'\;,
\end{equation}
the field which (at leading order) describes gravitational radiation
is given by
\begin{equation}
h_{ij} = \frac{G}{c^4}\frac{\ddot Q_{ij}}{r} \sim \frac{Gm^{\rm
quad}}{rc^2}\frac{v^2}{c^2}\;.
\end{equation}
In the final order of magnitude formula, the mass $m^{\rm quad}$ is
the portion of the source's mass that participates in quadrupolar
motions.  Notice that the dimensionful constant coupling the radiation
$h_{ij}$ to the source $Q_{ij}$ is incredibly small:
\begin{equation}
\frac{G}{c^4} = \frac{6.673 \times
10^{-8}\,\mbox{cm}^3\,\mbox{sec}^{-2}\,\mbox{gm}^{-1}}{(2.998 \times
10^{10}\,\mbox{cm/sec})^4}
= 8.260\times10^{-50}\,\frac{\mbox{sec}^2}{\mbox{gm}\,\mbox{cm}}\;.
\end{equation}
Overcoming this tiny coupling requires enormous masses and speeds
close to the speed of light.  GW sources thus tend to be compact
objects --- objects that are both massive and small enough that they
can whip around at relativistic speeds.  It also means that, once
generated, the waves barely interact with intervening matter.  This
unfortunately includes our detectors, so that detecting GWs is quite
an experimental challenge; but it also means that sources cannot be
obscured, as is often the case with electromagnetic sources.  Thus,
GWs present an opportunity to open a window directly onto the most
violent and dynamic processes in the universe.

In this article, we quickly survey some of the most important LISA GW
sources.  A handy way of organizing these sources is by the spectral
character of their waves (which in turn drives how we plan to analyze
LISA data): {\it Stochastic sources}, with a broadband, flat or widely
peaked spectrum; {\it monochromatic sources}, which radiate at nearly
a pure tone in their rest frame; {\it chirping sources}, like a pure
tone which quickly sweeps through the LISA band; and {\it really
complicated chirping sources} --- similar to chirping sources, but
with a particularly ornate character that merits special
consideration.

\section{Stochastic sources from cosmological backgrounds}

Stochastic GWs are ``random'' waves, arising from the superposition of
many discrete, uncorrelated sources.  They typically combine to
produce a broadband spectrum that is nearly flat, or weakly peaked
near some fiducial frequency.

Some of the most anticipated stochastic sources are cosmological in
origin.  By far the most eagerly sought waves are GWs arising from
primordial fluctuations in the universe's spacetime metric which have
been parametrically amplified by inflation.  In energy density units
(commonly used to parameterize these waves; see {\cite{allen96}} for a
very readable discussion), these waves are flat across an extremely
wide band, ranging from frequencies $f \sim 10^{-16}\,{\rm Hz}$ to
roughly $10^{10}\,{\rm Hz}$.  The lower end of this scale corresponds
to the inverse Hubble scale when the universe became matter dominated;
below this frequency, the spectrum grows as $f^{-2}$.  At the high
end, the spectrum is cut off by the (short but finite) time scale over
which inflation ends and the universe enters a hot,
radiation-dominated phase; see Fig.\ 12 of {\cite{allen96}} and
associated discussion.  What's particularly exciting about these waves
is that the level of this spectrum is set by --- and thus encodes ---
the potential which drives inflation.  Thus, measuring these waves
would {\it directly probe inflationary physics}.

Unfortunately, these waves have such small amplitudes that direct
detection with LISA (or any other near term GW experiment) is
completely out of reach.  LISA should be able to detect a background
at a level $\sim 10^{-11}$ of the universe's closure density; current
estimates suggest that this is at least 4 or 5 orders of magnitude
from the sensitivity needed to measure these waves.  Direct detection
at the relatively high frequencies of man-made facilities requires
finding a relatively quiet band (without too many ``foreground''
sources) and a much more sensitive instrument than LISA.  At the
lowest frequencies, detection is plausible in the next decade or so
through cosmic microwave background (CMB) studies.  The tensor nature
of GWs has a unique impact on the polarization of CMB photons
{\cite{zs97,kks97}}.  The signal is weak compared to other signals in
the CMB, and potentially contaminated by foreground effects.
Searching for it is nonetheless one of the most important directions
in CMB physics today.

Although inflationary waves are out of LISA's reach, other
cosmological backgrounds may not be.  Backgrounds are associated with
phase transitions that occur as the universe expanded and cooled.  For
example, when the mean temperature was roughly 1 TeV, the electroweak
interaction separated into electromagnetic and weak interactions.
This separation was not spatially homogeneous --- some regions
underwent this transition before others.  Different regions were at
different densities; boundaries between regions collided and
coalesced, producing a GW background {\cite{allen96,kmk02}}.  More
recently, there has been a flurry of work in superstring theory
suggesting that cosmic strings may have been produced in the early
universe and then expanded to cosmic sizes (see, for example
{\cite{p05}} for an overview).  A network of such objects could
produce a background accessible to LISA; see Hogan's contribution to
these proceedings {\cite{craig}} for detailed discussion.

Though a relatively speculative source, cosmological stochastic
backgrounds are easily searched for, and there is a tremendous amount
of discovery space.  The probability of payoff may be low, but the
potential reward for discovery is extraordinarily high.

\section{Periodic sources: Binary stars in the galaxy}

The Milky Way contains trillions of stars in binary systems; each is a
GW generator.  A small fraction --- tens of millions --- are compact
(mostly white dwarf binaries) and generate GWs in the LISA band.  GW
emission causes the binaries' stars to gradually spiral towards one
another, driving the frequency to slowly ``chirp'' upwards.  At the
masses and frequencies we're discussing here, the chirp is extremely
slow:
\begin{eqnarray}
\dot f &=& \frac{48}{5\pi}\mu M^{2/3}(2\pi f)^{11/3}
\label{eq:fdot}\\
&=& 9.2\times 10^{-18}\,{\rm
Hz/sec}\times\left(\frac{M}{1\,M_\odot}\right)^{5/3}
\left(\frac{f}{1\,{\rm mHz}}\right)^{11/3}\;.
\end{eqnarray}
We've specialized to equal masses in the last line, putting the
reduced mass $\mu = M/4$; we are also using units with $G = 1 = c$, so
that $M_\odot = 4.92\times10^{-6}\,{\rm sec}$.  The frequency of these
sources barely changes over a mission lifetime: $\dot f\times T_{\rm
mission}$ is much less than the rough binwidth, $\delta f \sim
(\mbox{a few})/T_{\rm mission}$, except on the high end of the band,
$f \gtrsim 0.01\,{\rm Hz}$.  These sources are thus mostly
monochromatic, with a scattering of slowly chirping ones at the high
end of the mass and frequency spectrum.

Such objects are extremely important for LISA science because they are
{\it guaranteed sources}.  Quite a few target binaries have already
been catalogued by x-ray and optical studies (cf.\ contribution to
this meeting by Gijs Nelemans {\cite{gijs}}, and Neleman's webpage
{\cite{gijsweb}}).  Indeed, population synthesis indicates there will
be so many binaries at low frequencies that they will form a confused
background --- so many binaries radiate in a single frequency bin that
they cannot be distinguished.  For studying certain sources (e.g.,
massive black hole binaries, discussed in the next section), this
background must actually be regarded as {\it noise}.  It is of course
{\it signal} to those interested in stellar populations!  For example,
Matt Benacquista and colleagues have shown that a galactic population
can be clearly distinguished from a population based in globular
clusters {\cite{bdl04}}, and Benacquista and Holley-Bockelman have
shown that the galactic scale height has important consequences for
the mean level and detailed characteristics of this background
{\cite{bhb06}}.  A wealth of data about the distribution of stars in
our galaxy is encoded in this ``noise''.

\section{Chirping sources: massive black hole coalescence}

Chirping sources can be thought of as a pure tone (like monochromatic
periodic sources), but with that tone rapidly sweeping through the
LISA band.  Consulting Eq.\ (\ref{eq:fdot}), we see that making a
binary sweep through the band requires large masses.  The most
important chirping LISA sources will be binaries in which both members
are massive black holes: With a total system mass of $10^4\,M_\odot -
10^7\,M_\odot$ and a mass ratio of $1/20 - 1$ or so, a binary
generates waves right in LISA's main band of sensitivity, and sweeps
across the band in a time ranging from a few months to a few years.

Binaries at these masses are created by the mergers of large
structures --- galaxies and protogalaxies, and the dark matter halos
which host them.  It has long been appreciated that the formation of
massive binary black holes as a ``side effect'' of hierarchical
structure formation could lead to healthy event rates.  Quoting from
Haehnelt's contribution to the Proceedings of the 2nd LISA Symposium
{\cite{mh98}}, ``even a pessimist who assumes a rather long QSO
lifetime and only one binary coalescence per newly-formed halo should
expect a couple of SMBH binary coalescences during the lifetime of
LISA while an optimistic might expect to see up to several hundred of
these exciting events.''

Recent work suggests the rate may be near the high end of this range,
though with most events at moderately high redshift.  One input to
this story is that it is now appreciated that almost all galaxies host
a massive black hole at their core.  It had long been argued that
accretion onto black holes must power the emission of quasars and
active galactic nuclei {\cite{lb69}}; a corollary is that the cores of
many galaxies should host massive black holes as ``quasar fossils''.
This expectation has been confirmed by observations which demonstrate
kinematically that the cores of galaxies with large central bulges
host a massive ($10^6-10^9\,M_\odot$) black hole.  Further, it has
been established that the properties of these holes are strongly
correlated with the properties of the galaxies that host them
{\cite{fm00,g00}}, indicating that the growth of black holes and their
host galaxies is tightly coupled.

Second, there is a consensus that galaxies grow hierarchically,
through repeated mergers of smaller galaxies as their host dark matter
halos repeatedly merge; see for example Hopkins et al.\
{\cite{hetal06}}.  Combining the apparent ubiquity of black holes in
galaxies with the tendency of galaxies to merge suggests that the
formation of massive binary black holes is fairly frequent, especially
at high redshift when mergers were common.  For example, Sesana et
al.\ {\cite{setal04}} calculate a total rate for LISA of about 60 --
70 events per year, with most ($\sim 50$) mergers occuring for a total
system mass smaller than $10^5\,M_\odot$ and for redshifts $z > 10$.
(Note that this mass is smaller than has been observed for any
galactic core black hole, but is a reasonable ``seed'' mass for a hole
that evolves to the masses observed in the local universe.)  About
$10$ mergers per year are expected to occur in the mass range
$10^5\,M_\odot \lesssim M \lesssim 10^6\,M_\odot$ and in the redshift
range $2 \lesssim z \lesssim 6$.  See Marta Volonteri's contribution
to this meeting for further details {\cite{marta}}.

\begin{figure}
  \includegraphics[height=0.3\textheight]{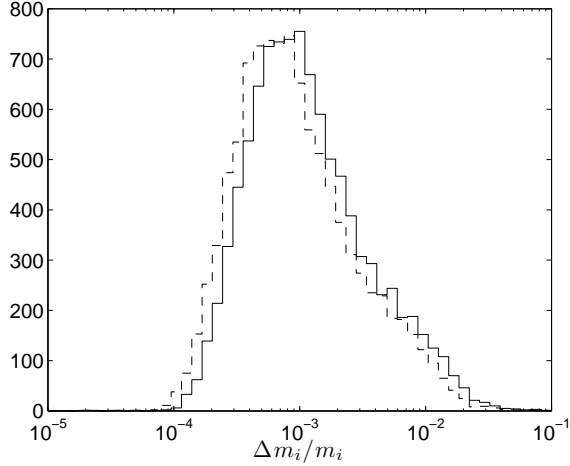}
  \caption{Distribution of measured mass errors from a Monte-Carlo
    simulation of $10^4$ binaries, randomly distributed on the sky,
    randomly oriented, and with random spins and spin orientations.
    All binaries are placed at redshift $z = 1$, and have masses $m_1
    = 10^6\,M_\odot$, $m_2 = 3\times10^5\,M_\odot$.  The solid line
    gives the distribution of errors in $m_1$, the dotted line gives
    the distribution in $m_2$.  Both distributions peak at a relative
    error of about $0.1\%$, and are almost entirely confined to errors
    smaller than $1\%$.}
  \label{fig:masses}
\end{figure}

\begin{figure}
  \includegraphics[height=0.3\textheight]{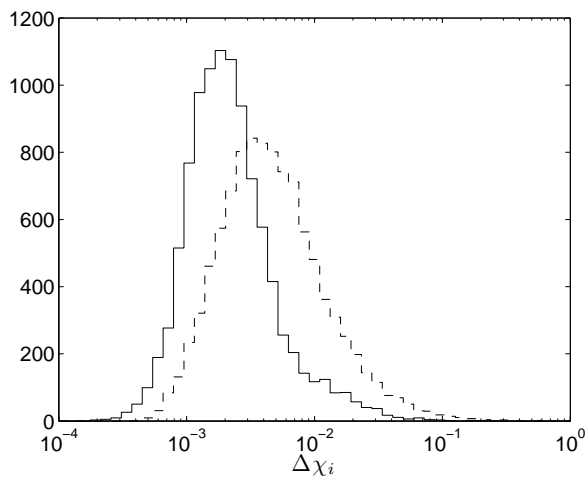}
  \caption{Same as Fig.\ {\ref{fig:masses}}, but now showing errors in
    of the Kerr spin parameter $\chi = |{\bf S}|/M^2$.  For most
    events in the distribution, the spins are determined to within
    $\delta\chi = 0.01$.  The spin of the larger black hole (solid
    line) is typically better measured than that of the smaller hole
    (dashed line), since the magnitude of its spin vector $|{\bf S}|$
    is larger and has a more important impact on the waveform.}
  \label{fig:spins}
\end{figure}

In the months to a year or so that the binary radiates in LISA's most
sensitive band, many thousands to tens of thousands of wave cycles
accumulate.  Even for high redshift sources, these signals can be
detected with very high signal-to-noise ratio (of order hundreds after
convolving with a model template).  By tracking these many wave cycle
at such high signal-to-noise, it will be possible to determine binary
parameters with exquisite accuracy.  Figure {\ref{fig:masses}} (taken
from Ref.\ {\cite{lh06}}) shows how well masses can be determined ---
for this example, both black hole masses are determined with relative
errors that are typically much less than $1\%$.  This is because the
masses strongly influence the rate at which the orbital phase evolves.
In addition, the spins of the binary's member holes impact the
waveform through precessional effects; by modeling those precessions,
the black hole spins can be determined.  Figure {\ref{fig:spins}}
(also taken from Ref.\ {\cite{lh06}}) shows that, for the same
distribution of binaries as in Fig.\ {\ref{fig:masses}}, the Kerr spin
parameter $\chi = |{\bf S}|/M^2$ can typically be determined to within
$0.01$.  {\it LISA will be a tool for the precision measurement of the
cosmic growth of black hole masses and spins.}  This will be a
beautiful probe of the coevolution of black holes and galaxies.

``Extrinsic'' source parameters --- those having to do with its
location and orientation relative to the detector --- are also
determined by measuring its GWs.  What is particularly interesting is
that GWs {\it directly encode the luminosity distance to the source}.
In essence, inspiralling binaries are a standard candle in which the
standardization is provided by general relativity.  As shown in Ref.\
{\cite{lh06}}, source distances can be measured to within a few
percent, even for fairly high redshift.

The position of a binary on the sky is not determined terribly well by
astronomy standards, but is determined well enough that planned large
scale surveys may be able to search for ``electromagnetic
counterparts'' to merging binary black holes.  At low redshift,
sources are typically localized to an ellipse whose major axis is
roughly $10 - \mbox{a few 10s}$ of arcminutes, and whose minor axis is
typically a factor of $2 - 5$ smaller.  At higher redshift, this
``pixel'' bloats by a factor of a few in each direction, so that a
field of a few square degrees may need to be searched {\cite{lh06}}.
Hopefully, some kind of unique signature will be associated with the
merger event, such as the onset of AGN or quasar activity.  If this is
the case, then the chance of associating a merger event with an
electromagnetic counterpart is quite high {\cite{bence}}.

Massive binary black hole inspirals provide perhaps the most
broad-reaching, exciting LISA science.  These sources can be measured
to large redshifts, and can be measured well enough to determine the
parameters of the black holes with high precision.  Measuring these
waves will make LISA an instrument for untangling the growth of black
holes and tracking cosmic structure evolution.

\section{{\it Really complicated} chirping sources: Extreme mass
ratio inspirals}

Extreme mass ratio inspirals (EMRIs) are binary systems in which one
member is a massive, galactic core black hole, and the other is a
stellar mass compact object.  Such binaries are created by scattering
processes in the cores of galaxies --- the smaller member of the
binary is scattered by multibody interactions onto a highly eccentric,
strong-field orbit of the massive black hole.  Current estimates
suggest that hundreds of these sources may produce GWs accessible to
LISA each year.  Further details can be found in Clovis Hopman's
contribution to these proceedings {\cite{clovis}}.

Once scattered onto this orbit, the EMRI becomes a strong GW radiator.
The small body executes $10^4 - 10^5$ orbits as it spirals in.  These
orbits are quite ornate.  A strong-field black hole orbit has three
orbital periods: $T_\phi$, describing motion in the axial direction;
$T_\theta$, describing poloidal oscillations; and $T_r$, describing
radial oscillations.  These periods coincide in the Newtonian limit
(mean orbit radius $r \gg M$), but differ quite strongly in the strong
field, with $T_r > T_\theta > T_\phi$.  The mismatch between for
example $T_r$ and $T_\phi$ means that an orbiting body can appear to
``whirl'' around the black hole many times as at ``zooms'' in from
apoapsis to periapsis and back.  This ``zoom-whirl'' character leaves
a distinctive stamp on the binary's GWs {\cite{gk02}}.

The character of these modulations and the manner in which they evolve
as the small body spirals in encodes a great deal of information about
the binary's spacetime.  {\it If} it is possible to coherently track
the inspiral over those $10^4 - 10^5$ orbits, it should be possible to
determine the character of the spacetime with very high precision.  It
should be emphasized that building models capable of tracking the
inspiral is quite a difficult task; see the author's other
contribution to these proceedings for further discussion
{\cite{me_emri}}.

A relatively prosaic application of these measurements will be to
determine the mass and spins of quiescent black holes.  Using a
simplified model for EMRI waves, Barack and Cutler have shown that
LISA will be able to determine massive black hole masses and spins
with an accuracy of about a part in $10^4$ {\cite{bc04}}.  Such
measurements will make it possible to precisely survey the
distribution of black hole spins and masses in the nearby\footnote{In
contrast to massive black hole coalescences, most EMRI events will be
relatively close by --- $z \lesssim 1$.} universe.  More
fundamentally, one can analyze these sources without assuming {\it a
priori} that the massive object is described by the Kerr spacetime.
The measurement then tests how well the Kerr hypothesis fits the data.
More discussion of how this test can be done is given in the author's
other contribution to these proceedings {\cite{me_emri}}.

\section{Summary}

LISA is an astrophysics mission that will make possible unprecented
{\it precision} measurements of dynamic, strong-gravity phenomena.
Perhaps most exciting will be the ability to track the mergers of
massive black holes, precisely weighing their masses and spins.  Such
measurements will directly probe the early growth of black holes,
opening a new window onto structure growth.

\begin{theacknowledgments}
Support for this work is provided by NASA Grants NAG5-12906 and
NNG05G1056; work on gravitational-wave science generally is supported
by NSF Grants PHY-0244424 and PHY-0449884.  We also gratefully
acknowledge support from MIT's Class of 1956 Career Development Fund.
\end{theacknowledgments}

\bibliographystyle{aipprocl}

\end{document}